\documentclass[aps,prd,twocolumn,nofootinbib,preprintnumbers]{revtex4}
\usepackage[dvips]{graphicx}
\usepackage{bm,latexsym,amsmath,amssymb,amsfonts,mathrsfs}
\usepackage{color}
\input{colordvi.tex}

\newcommand{\del}{\partial}

\begin{document}

\title{Supersymmetric DBI inflation}

\author{Shin Sasaki}
\email[Email: ]{shin-s"at"kitasato-u.ac.jp}
\affiliation{%
Department of Physics, Kitasato University, Sagamihara 228-8555, Japan}

\author{Masahide Yamaguchi}
\email[Email: ]{gucci"at"phys.titech.ac.jp}
\affiliation{Department of Physics, Tokyo Institute of Technology, Tokyo 152-8551, Japan}

\author{Daisuke Yokoyama}
\email[Email: ]{d.yokoyama"at"th.phys.titech.ac.jp}
\affiliation{Department of Physics, Tokyo Institute of Technology, Tokyo 152-8551, Japan}

\begin{abstract}
We discuss a supersymmetric version of DBI (Dirac-Born-Infeld)
inflation, which is a typical inflation model in string cosmology.  The
supersymmetric DBI action together with a superpotential always leads to
correction terms associated with the potential into the kinetic term,
which drastically change the dynamics of DBI inflation.  We find two
significant features of supersymmetric DBI inflation.  The first one is
that ultra-relativistic motion is prohibited to cause inflation, which
leads to order of unity sound velocity squared and hence small
non-Gaussianities of primordial curvature perturbations. The second one
is that the relation between the tensor-to-scalar ratio and the field
variation is modified. Then, significant tensor-to-scalar ratio $r
\gtrsim 0.01$ is possible because the variation of the canonically
normalized inflaton can be beyond the reduced Planck scale. These new
features are in sharp contrast with those of the standard
non-supersymmetric DBI inflation and hence have a lot of interest
implications on upcoming observations of cosmic microwave background
(CMB) anisotropies by the Planck satellite as well as direct detection
experiments of gravitational waves like DECIGO and BBO.
\end{abstract}

\preprint{TIT/HEP-617}

\maketitle

Recent observations of CMB anisotropies like the Wilkinson Microwave
Anisotropy Probe (WMAP) satellite strongly suggest the presence of
accelerated expansion called inflation in the early Universe
\cite{Komatsu:2010fb}. During inflation, primordial curvature
\cite{Guth:1982ec} and tensor \cite{Starobinsky:1979ty} perturbations
are generated and stretched out to the cosmological scales, which become
seeds of the large scale structure formation and the CMB
anisotropies. The properties of primordial curvature fluctuations are
well known and are almost scale-invariant, adiabatic, and
Gaussian. Though tensor perturbations have not yet been found
unfortunately, they are expected to be detectable in upcoming CMB
experiments like the Planck \cite{Planck:2006aa} and the CMBPol
\cite{Baumann:2008aq}. However, we do not know the origin of the
inflaton at all, except that it is an effective scalar field. (See
Refs. \cite{Mazumdar:2010sa} for recent review of inflation model
building.)

String theory is the most powerful candidate to unify all of the
fundamental interactions. Then, it is natural to pursue the candidate of
an inflaton in string theory. In fact, inflation models in the brane
setting were proposed \cite{Dvali:1998pa,Kachru:2003sx} and have been
investigated intensively. Among them, a particularly interesting class
of inflation models is DBI inflation \cite{Silverstein:2003hf}, which is
associated with the relativistic motion of a D-brane in the warped flux
compactification. This model has distinctive predictions for primordial
perturbations: (i) it can naturally generate large non-Gaussianities of
primordial curvature perturbations thanks to the ultra-relativistic
motion \cite{Silverstein:2003hf}, (ii) it is quite difficult to produce
detectable tensor perturbations because the maximal field variation of
the inflaton is constrained to be less than the reduced Planck scale
$M_{\rm pl}$ \cite{Baumann:2006cd}. Current observations like the WMAP
satellite are precise enough to rule out simple UV models of DBI
inflation \cite{Bean:2007hc} though more elaborated models are still
compatible with the present observations
\cite{Chen:2005ad,Langlois:2008wt}.

Almost all of these studies of DBI inflation, however, have been based on the
non-supersymmetric
setup. Supersymmetry is one of
the most promising solutions to the hierarchy problem of the Standard
Model as well as the unification of the fundamental interactions. Once a
probe D-brane is placed on supersymmetric backgrounds, one expects that
the world-volume effective theory of the probe brane becomes
supersymmetric. Therefore, it is quite important to consider DBI
inflation in the supersymmetric framework. Recently, some attempts to
supersymmetrize non-canonical kinetic terms have been done
\cite{Khoury:2010gb,Baumann:2011nk}. However, in order to incorporate a
potential term, one needs to introduce a superpotential and solve the
equation of motion for the auxiliary field consistently. This is a
difficult task when the non-canonical kinetic terms are present.

In this Letter, we discuss the supersymmetric version of DBI
inflation. First of all, the supersymmetric DBI action with the
superpotential is studied. By solving the equation of motion for the
auxiliary field consistently, we show that correction terms associated
with the potential always appear in the kinetic term, which drastically
changes the dynamics of DBI inflation. Then, using the newly obtained
action, we investigate the dynamics of supersymmetric DBI inflation in
detail. It is revealed that the predictions of primordial perturbations
are completely different from those of non-supersymmetric DBI inflation,
which may require us to reanalyze all of the DBI inflation models
including elaborated models in preparation for the upcoming experiments.

Now let us begin with the supersymmetric DBI action in the warped
throat.  Assuming that a probe D3-brane is moving in the supersymmetric
ten-dimensional geometry of the form
\begin{equation}
d s_{10}^2 = H^{-\frac{1}{2}} (y) d s_4^2 + H^{\frac{1}{2}} (y) d s_6^2,
\label{eq:geometry}
\end{equation}
where $ds_4^2$, $ds_6^2$ are four-dimensional spacetime and
six-dimensional internal space respectively, the supersymmetric DBI
Lagrangian in the flat spacetime \cite{Rocek:1997hi} is generalized as
follows:
%
\begin{widetext}
\begin{equation}
  {\cal L}_{\rm DBI} = \int d^4 \theta \left[
      \Phi \Phi^{\dagger} 
      +
\frac{1}{16 T}
\left(D^{\alpha}\Phi D_{\alpha}\Phi \right)
       \left(\overline{D}^{\dot{\alpha}}\Phi^{\dagger}
	\overline{D}_{\dot{\alpha}}\Phi^{\dagger} \right)  
       \frac{1}{1+A+\sqrt{\left(1+A\right)^2-B}}
  \right],
\label{eq:SDBI}
\end{equation}
\end{widetext}
where we have employed the static gauge and 
normalized the D3-brane tension and the string slope parameter $2\pi \alpha'$ to
unity.  The chiral and anti-chiral superfields are denoted by $\Phi$ and
$\Phi^{\dagger}$, $D_{\alpha}$ and $\overline{D}_{\dot{\alpha}}$ are the
supercovariant derivatives, $T = T\left(\Phi, \Phi^{\dagger}\right)$ is
a function of $\Phi, \Phi^{\dagger}$ corresponding to the warp factor
$T = H^{-1}$, and $A$, $B$ are given by
\begin{equation}
  A \equiv \frac{\del_{\mu}\Phi \del^{\mu}\Phi^{\dagger}}{T}, \quad 
  B \equiv \frac{\del_{\mu}\Phi \del^{\mu}\Phi 
                 \del_{\nu}\Phi^{\dagger} \del^{\nu}\Phi^{\dagger}}
{T^2}.
\end{equation} 
Here we have turned on one complex scalar field associated with two
independent fluctuations along the throat direction $y$. In order to
incorporate the potential, we add the superpotential term to the
Lagrangian,
\begin{equation}
  {\cal L}_{\rm pot} =  \int d^2 \theta~W\left(\Phi\right) + {\rm h.c.}.
\end{equation}
Several kinds of superpotentials are induced by the background fluxes.
For example, we can introduce the superpotential of a mass term $W =
\frac12 m \Phi^2$ on the D3-brane in the presence of a constant
Ramond-Ramond 3-form background \cite{Ito:2007hy, Billo:2008sp}. 

The component Lagrangian is
\begin{widetext}
\begin{eqnarray}
\mathcal{L} &=& - T
\sqrt{
1 + 2 T^{-1} \partial_{\mu} \varphi \partial^{\mu} \overline{\varphi}
+ T^{-2} (\partial_{\mu} \varphi \partial^{\mu} \overline{\varphi})^2
- T^{-2} (\partial_{\mu} \varphi \partial^{\mu} \varphi)
(\partial_{\nu} \overline{\varphi} \partial^{\nu} \overline{\varphi})
} + T
\nonumber \\
& & + \overline{F} F + \frac{\partial W}{\partial \varphi} F 
+ \frac{\partial \overline{W}}{\partial \overline{\varphi}} \overline F
+ G(\varphi) (- 2 F \overline{F} \partial_{\mu} \varphi \partial^{\mu} \overline{\varphi}
+ F^2 \overline{F}^2),
\end{eqnarray}
\end{widetext}
where we have dropped the fermions since they do not contribute to the
dynamics of the inflation. The function $G(\varphi)$ is defined as
\begin{equation}
  G(\varphi) = \frac{1}{T}
   \frac{1}{1+A+\sqrt{\left(1+A\right)^2-B}},
\end{equation}
with the replacement of $\Phi$ by $\varphi$ in $A$ and $B$. 

The 
Lagrangian for
the scalar component $\varphi$ 
in the chiral superfield $\Phi$
is obtained by solving the equation of motion for the auxiliary field
$F$ in $\Phi$. This is in general a simultaneous equation for $F$ and $\bar{F}$.
After eliminating $\bar{F}$, we find the equation for $F$ is given by  
\begin{equation}
  2G(\varphi) \frac{\del W}{\del \varphi}F^3 
  +\frac{\del \overline{W}}{\del \overline{\varphi}}
   \left(1-2G(\varphi)\del_{\mu} \varphi \del^{\mu}\overline{\varphi}
   \right)F
  + \left( \frac{\del \overline{W}}{\del \overline{\varphi}} \right)^2
  = 0.
\label{eq:F}
\end{equation}
%
%
%
Unlike
the standard (quasi-)canonical case, a salient feature of the
supersymmetric DBI model is that the equation for $F$ is {\it cubic} and
can be solved analytically by Cardano's method,
\footnote{
In the case where the fermions are present, 
a perturbative solution to the equation for $F$ 
was discussed in \cite{Khoury:2010gb}.
}
\begin{eqnarray} 
  F &=& \omega^{k} \sqrt[3]{-\frac{q}{2}
      +\sqrt{\left(\frac{q}{2}\right)^2+\left(\frac{p}{3}\right)^3}} \nonumber\\
    && +\omega^{3-k} \sqrt[3]{-\frac{q}{2}
      -\sqrt{\left(\frac{q}{2}\right)^2+\left(\frac{p}{3}\right)^3}}.
\label{eq:aux_sol}
\end{eqnarray}
Here $\omega$ is the complex cubic root, $k = 0,1,2$, and $p$ and $q$
are given by
\begin{eqnarray}
  p &=& \left( \frac{\del W}{\del \varphi} \right)^{-1}
      \frac{\del \overline{W}}{\del \overline{\varphi}}
      \frac{1-2G \del_{\mu} \varphi \del^{\mu}\overline{\varphi}}
           {2G}, \nonumber \\
  q &=& \frac{1}{2G} \left( \frac{\del W}{\del \varphi} \right)^{-1}\left( \frac{\del \overline{W}}{\del
                    \overline{\varphi}} \right)^2.
\label{eq:pq}
\end{eqnarray}
We note that if $W=0$, the unique solution is given by $F=0$ and the
bosonic part of the DBI Lagrangian \eqref{eq:SDBI} is not changed
compared with the non-supersymmetric case. A remarkable fact in the case
$W \not=0$ is that there are three different on-shell actions associated
with the $k=0,1,2$ solutions in Eq. \eqref{eq:aux_sol}. In the following, we
concentrate on the $k=0$ branch since it is continuously connected to
the ordinary solution $F = - \del \overline{W}/\del \bar{\varphi}$ in
the canonical limit. The other solutions with $k=1,2$ do not have any
definite limit and will yield essentially inequivalent theories.

Now we denote the phase factor of $\partial W/\partial \varphi$ as
$\alpha$.  Since the functions $A$, $B$ and $G$ are real, the phase of
$p$ and $q$ in the solutions \eqref{eq:pq} are $- 2 \alpha$ and $- 3
\alpha$ respectively.  Then from the $k=0$ solution in
\eqref{eq:aux_sol}, the phase of $F$ is given by $\pi -\alpha$.  As a
result, the phase factor of the product of $\partial W/\partial \varphi$
and $F$ becomes $\pi$ and does not depend on $\alpha$ in the on-shell
Lagrangian.  Therefore only the absolute values of $\partial W/\partial
\varphi$ and $F$ contribute to the Lagrangian.

We further impose the global $U(1)_R$ symmetry on the superpotential
$W(\Phi)$ and the warp factor $T$. This is always possible when the
geometry \eqref{eq:geometry} has a $U(1)$ isometry in the $y$
direction. A typical example of this kind of geometry is the near
horizon limit of $N$ coincident D3-branes
\cite{Silverstein:2003hf}. Since the supersymmetric DBI Lagrangian given
in Eq. (\ref{eq:SDBI}) is invariant under the $U(1)_R$ symmetry, 
the dynamics of the scalar field $\varphi$ depends only on its radial
component $f$.
In this case, $f$ is identified with the fluctuation along the
radial direction in $AdS_5 \times S^5$. 

Under these circumstances, the full on-shell action for the scalar field
$f$ in curved spacetime is given by
\begin{eqnarray}
  S &=& \int d^4 x \sqrt{-g} 
    \left( \frac12 M_{\rm pl}^2 R +{\cal L}_{f} \right), \\
  && {\cal L}_f = {\cal L}_{\rm DBI} +{\cal L}_{\rm aux}, \\
  && {\cal L}_{\rm DBI} = T \frac{\gamma-1}{\gamma}, 
\label{eq:DBIreal} \\
  && {\cal L}_{\rm aux} =
      F^2-2\sqrt{2}~\frac{dW}{df}F
      + G F^2 \left( 2X + F^2 \right),
\label{eq:auxreal}
\end{eqnarray}
where $F$ and $\partial W/\partial f$ are real and positive, and
%
\begin{equation}
  X \equiv -\frac12 \del_{\mu} f \del^{\mu} f, \,\, 
  \gamma = \frac{1}{\sqrt{1-2\frac{X}{T}}},\,\, 
  G = \frac{1}{T}\frac{2\gamma^2}{(1+\gamma)^2}.
\end{equation}
The equation for the auxiliary field
can be rewritten as
\begin{equation}
  2GF^3 + \left(1+2GX \right)F - \sqrt{2}~\frac{dW}{df} = 0.
  \label{eq:Freal}
\end{equation}
Although we have the analytic solutions to the above equation, since its
complexity would make it difficult to capture the essence of the
physical properties,
we look for approximate solutions under the assumption that one term in
the left-hand side of Eq. \eqref{eq:Freal} is subdominant. This will
provide a valuable intuition for the clear characteristics of our model.
Later, we will mention the case that all the terms in
Eq. \eqref{eq:Freal} are comparable.

Case (i) : subdominance of the first term. The auxiliary field $F$ is
given by
\begin{equation}
   F \simeq \sqrt{2}~\frac{\gamma+1}{3\gamma-1}\frac{dW}{df}.
\end{equation}
The condition that the first term is negligible is satisfied for 
\begin{equation}
  \frac{8\gamma^2(\gamma+1)}{(3\gamma-1)^3} 
    \frac{1}{T}\left(\frac{dW}{df}\right)^2 \ll 1.
  \label{eq:cond1}
\end{equation}
Since the prefactor in the left-hand side of the above inequality is of
the order unity for $\gamma \ge 1$, $\left(dW/df\right)^2 \ll T$. Under
this condition, the Lagrangian ${\cal L}_f$ for the scalar field $f$ is
dominated by the DBI kinetic term ${\cal L}_{\rm DBI}$ only except
the case $\gamma - 1 \ll 1$ when the DBI kinetic term is significantly
suppressed. Hence, only inflation with a usual (almost) canonical
kinetic term can happen. 
In this case, large tensor
perturbations are prohibited due to the Lyth bound and the constrained
field variation.

Case (ii) : subdominance of the last term. The solution is obtained
by taking the limit $q \to 0$ in Eq. \eqref{eq:aux_sol} and is found to
be $F=0$, which leads to no potential and hence no inflation.


Case (iii) : subdominance of the middle term. 
The solution is given by taking the limit $p \to 0$ in Eq. \eqref{eq:aux_sol}. 
We obtain
\begin{equation}
   F \simeq \frac{1}{\sqrt{2}}
            \left( \frac{1+\gamma}{\gamma} \right)^{\frac23}
            \left( T \frac{dW}{df} \right)^{\frac13}.
\end{equation}
The subdominance of the middle term is satisfied for
\begin{equation}
  \frac{8\gamma^2(\gamma+1)}{(3\gamma-1)^3} 
    \frac{1}{T}\left(\frac{dW}{df}\right)^2 \gg 1.
  \label{eq:cond2}
\end{equation}
Note that this condition is just the opposite inequality of
Eq. (\ref{eq:cond1}) and equivalent to $\left(dW/df\right)^2 \gg T$.
Substituting this solution of $F$ in Eq. (\ref{eq:auxreal})
yields
\begin{equation}
  {\cal L}_{\rm aux} = - \frac{1}{2^{\frac23}} 
     \left( \frac{\gamma+1}{\gamma} \right)^{\frac23} V(f),
\end{equation}
where the potential $V(f)$ is defined so that ${\cal L}_{\rm aux}
\rightarrow - V(f)$ for $\gamma \rightarrow 1$ (i.e. no kinetic term
limit $X \to 0$),
\begin{equation}
  V(f) \equiv \left(\frac{27T}{2}\right)^{\frac13} 
              \left(\frac{dW}{df}\right)^{\frac43}.
\end{equation}
The condition (\ref{eq:cond2}) can be recast into $V \gg T$.  Note that
this is not a sufficient condition for inflation because the kinetic
terms ($\gamma$) depending on the potential appears in ${\cal L}_{\rm
aux}$.  In fact, the slow-roll parameter $\epsilon$ is given by
\begin{equation}
  \epsilon = - \frac{\dot{H}}{H^2} \simeq \frac{3(\gamma-1)}{2\gamma+1},
\end{equation}
where we have used $V \gg T$. Thus, inflation can happen only for
$\gamma \simeq 1$, that is, the ultra-relativistic motion of the D-brane
is prohibited in the supersymmetric DBI inflation, which is in marked
contrast to the standard non-supersymmetric case.  For k-inflation type
Lagrangian (${\cal L}_f=K(f,X)$) \cite{ArmendarizPicon:1999rj} including
the DBI inflation as a special case, the non-Gaussianities of the
curvature perturbations are enhanced by $1/c_s^2$
\cite{Seery:2005wm}. Then, the standard non-supersymmetric DBI inflation
predicts large non-Gaussianities for ultra-relativistic motion because
of $c_s^2 = 1/\gamma^2$ \cite{Silverstein:2003hf}. On the other hand, in
our case, the sound velocity squared are estimated as
\begin{equation}
  c_s^2 \simeq 3/(3\gamma^2+\gamma-1) \simeq 1,
\end{equation}
for $V \gg T$ and $\gamma \simeq 1$. Thus, $c_s^2$ becomes almost unity,
and hence negligible non-Gaussianity is predicted for the supersymmetric
DBI inflation. Next, we discuss tensor perturbations and comment on the
generalized Lyth bound \cite{Baumann:2006cd,Lyth:1996im}. The field
variation of $f$ can be related to the e-folding number $N$ for ${\cal
L}_f=K(f,X)$ as,
\begin{equation}
  \frac{df}{M_{\rm pl}} = \sqrt{\frac{r}{8 c_s K_X}} dN,
\end{equation}
where $r$ is the tensor-to-scalar ratio and $K_X$ is the partial
derivative of $K$ with respect to $X$. Here, you should notice that $c_s
K_X = 1$ both for the canonical kinetic term ($c_s = K_X=1$) and
for the standard DBI case $c_s^{-1} = \gamma = K_X$, which leads
to the so-called Lyth bound, namely, significant tensor-to-scalar ratio
$r \gtrsim 0.01$ is possible only for $\Delta f \gtrsim M_{\rm
pl}$. However, the relation $c_s K_X = 1$ does not hold true in our
case. Instead, the following relation is obtained for $\gamma \sim 1$
and $V \gg T$,
\begin{equation}
   c_s K_X \sim \frac{V}{3T} \gg 1. 
\end{equation}
Therefore, the tensor-to-scalar ratio $r$ is enhanced by the factor $c_s
K_X$ in comparison to the standard non-supersymmetric DBI inflation,
which leads to significant tensor-to-scalar ratio $r \gtrsim 0.01$ even
for apparent sub-Planck variation of the field. This can be easily
understood by expanding the Lagrangian around $\gamma = 1$ and taking
the leading terms for $V \gg T$,
\begin{equation}
  {\cal L}_{f} \simeq \frac{V}{3T}X - V.
\end{equation}
Thus, the kinetic term is enhanced by $V/{3T}$. If we take the canonical
kinetic term by redefining the field $f$ as $f_{\rm can} \sim f
\sqrt{V/(3T)}$, the Lyth bound applies for $f_{\rm can}$.
\footnote{Even in our case, Planck-suppressed operators for the
canonically normalized field $f_{\rm can}$ must be controlled to
guarantee large tensor perturbations \cite{Baumann:2011ws}. One of such
methods is to introduce (approximate) shift symmetry
\cite{Kawasaki:2000yn}. It is manifest from the chiralities of $\Phi$
and $\Phi^{\dagger}$ that our DBI action given in Eq. \eqref{eq:SDBI}
can be easily modified to respect it approximately. However, it should
be notice that we have to abandon a global $U(1)_R$ symmetry in this
case, though the analysis runs almost parallel.} Therefore, the
observable tensor perturbations are predicted because the variation of
the canonically normalized inflaton $f_{\rm can}$ can be beyond the
reduced Planck scale.
Finally, we would like to mention the case that all of the terms in the
left-hand side of Eq. \eqref{eq:Freal} are comparable. Under this
condition, $V \sim (dW/df)^2$ and $T$ are comparable. Then, by comparing
the kinetic part (\ref{eq:DBIreal}) and the auxiliary part
(\ref{eq:auxreal}) in the Lagrangian, it is easy to verify that
inflation is possible only for $\gamma - 1 \ll 1$ in this case as well.

In summary, we have discussed the supersymmetric DBI inflation. In order
to accommodate the potential term in addition to the DBI kinetic term
consistently, the equation of motion for the auxiliary field $F$ is
derived and solved. Inserting its solution into the Lagrangian, we
obtain the effective Lagrangian for the supersymmetric DBI inflation, in
which the kinetic term related to the potential always appears in
addition to the DBI kinetic term. We find that ultra-relativistic motion
of the D-brane is forbidden to cause inflation, which has the
significant implications on the prediction of the primordial
perturbations. Firstly, the non-Gaussianities of the primordial
curvature perturbations are negligible because the sound velocity
squared are almost unity. Second, the significant tensor-to-scalar ratio
is possible in our model, especially in Case (iii), because of the
enhancement of the kinetic term. These two features are totally
different from those of the standard non-supersymmetric DBI
inflation. Provided that our model be realized, upcoming observations
such as Planck and CMBPol experiments will detect such tensor
perturbations though the non-Gaussianities of the curvature
perturbations will, unfortunately, not be observed.

These new predictions are based on the fact that one always encounters
kinetic (derivative) terms accompanied by the potential in
supersymmetric models with non-canonical kinetic terms. This feature is
not confined to DBI inflation but quite generic to non-trivial kinetic
terms appearing in inflation models such as k-inflation
\cite{ArmendarizPicon:1999rj} and G-inflation \cite{Kobayashi:2010cm},
which must also be supersymmetrized once supersymmetry would be found as
fundamental symmetry. For example, similar structures, such as a cubic
equation of the auxiliary field and potential-induced kinetic terms,
appear in the k-inflation models with superpotentials. We will discuss
elsewhere the supersymmetrization of these models and its implications
for cosmology by solving the equation for an auxiliary field adequately.

\paragraph*{Acknowledgments}

We would like to thank Tsutomu Kobayashi, Shinji Mukohyama, and Jun'ichi
Yokoyama for useful comments. This work is supported in part by the
Grant-in-Aid for Scientific Research No.~21740187 (M.Y.) and the
Grant-in-Aid for Scientific Research on Innovative Areas No.~24111706
(M.Y.).  D.Y. acknowledges the financial support from the Global Center
of Excellence Program by MEXT, Japan through the ``Nanoscience and
Quantum Physics'' Project of the Tokyo Institute of Technology.

\newpage


\end{document}